\documentclass[aps,preprint,pra,graphicx]{revtex4-2}
\usepackage{graphicx}
\usepackage{amsmath,amssymb}
\usepackage{scalerel,stackengine}

\stackMath
\newcommand\reallywidehat[1]{%
\savestack{\tmpbox}{\stretchto{%
  \scaleto{%
    \scalerel*[\widthof{\ensuremath{#1}}]{\kern-.6pt\bigwedge\kern-.6pt}%
    {\rule[-\textheight/2]{1ex}{\textheight}}
  }{\textheight}%
}{0.5ex}}%
\stackon[1pt]{#1}{\tmpbox}%
}

\makeatletter
\newcommand{\dalembertian}{\mathop{\mathpalette\dalembertian@\relax}}
\newcommand{\dalembertian@}[2]{%
  \begingroup
  \sbox\z@{$\m@th#1\square$}%
  \dimen0=\fontdimen8
    \ifx#1\displaystyle\textfont\else
    \ifx#1\textstyle\textfont\else
    \ifx#1\scriptstyle\scriptfont\else
    \scriptscriptfont\fi\fi\fi3
  \makebox[\wd\z@]{%
    \hbox to \ht\z@{%
      \vrule width \dimen0
      \kern-\dimen0
      \vbox to \ht\z@{
        \hrule height \dimen0 width \ht\z@
        \vss
        \hrule height 2\dimen0
      }%
      \kern-2.5\dimen0
      \vrule width 2.5\dimen0
    }%
  }%
  \endgroup
}
\makeatother

\begin{document}

\title{From Faraday and Maxwell to Quantum Physics.
The later story of the Electromagnetic Vector Potential.}

\author{Tuck Choy and Miguel Ortu\~no}
\email{tuckvk3cca@gmail.com}
\affiliation{Departmento de Fisica-CIOyN, Universidad de Murcia, Murcia 30071, Spain.}


\begin{abstract}

With the advent of quantum mechanics by Heisenberg in 1925 exactly a century ago, the quantization of the electromagnetic field became an important goal for our founding fathers, whom we are here to celebrate.  It was realized very soon that a consistent picture of quantum electrodynamics (QED) requires the quantization of not just the electromagnetic field ${\bf A}({\bf r},t)$, but the electron field ${\bf \psi}({\bf r},t)$ as well.  The electron field is now a Dirac spinor field and it becomes a major partner for the vector potential.  Together these two fields form a duet which yearns for a unification which is not often emphasized in text books, but no one knows how to do this yet. The underlying structure appears to be a super-symmetric one where bosons and fermions live comfortably as a single field that could by some yet unknown process produce QED. This area is still at the forefront of current research with names like Super-symmetric Quantum Electrodynamics (SQED), strings and others. Ironically for a variety of reasons, the number of fields in SQED has to be increased and not decreased, defying the objective of an economical unification.  Furthermore it is now known that Maxwell's theory is incomplete and that there are situations due to boundaries or topology that demand the vector potential must itself be gauge invariant.  This is another topic of current research which we can only briefly survey in this paper. To keep one's feet to the ground, we will show some novel toy models such as those proposed by Jennison (1979) which will be used to illustrate how one might ``create" an electron from just pure light. While crude, these models should be looked at in the spirit of Maxwell's original ``mechanical models" for light which despite their crudeness, were able to guide Maxwell to propose the displacement current, central to his electromagnetic theory, see Basil Mahon's
paper \cite{BasilMahon1}. We will also show that it is unexpectedly easy to operate in the QED regime, (no big particle accelerators needed) with something as common as an everyday mobile phone. Physics students may now do experiments in QED even from their home kitchen table! We shall recall attention to some of Heisenberg, Dirac and Wigner's insightful remarks in 1972 on the occasion of Dirac's 70th birthday symposium in Trieste, and end with some speculations about gravity waves and quantum mechanics.

(This paper is dedicated to the celebration of 100 years of quantum mechanics, on the anniversary of Heisenberg's founding paper on the subject in July 1925, delivered by Miguel Ortu\~no at the IQSA 2025 Conference at Tropea in Calabria, Italy from 30 June to 4th July 2025. The proceedings will be published as a celebratory volume by World Scientific Publications, Singapore in 2026).

\end{abstract}

\maketitle

\begin{quote}

``All these fifty years of conscious brooding have brought me no nearer to the answer to the question, `What are light quanta?' Nowadays every Tom, Dick and Harry thinks he knows it, but he is mistaken Albert Einstein \cite{Einstein1}."
\end{quote}

\section{Introduction}
\label{Introduction}
Let us begin by reminding the reader of the above quote from the Einstein who first discovered the `photon' in 1905.  Einstein was known to have gone to Westminster Abbey in June 1921 for 
the first time to lay a wreath on the tomb stone of Isaac Newton. Reporters surrounding him asked if he thought he was the successor to Newton.  Einstein replied that
 ``That statement is not quite right, I stood on Maxwell's shoulders".  Most reporters at that time we would guess did not know who James Clerk Maxwell was.
  Maxwell would have pointed out that he in turn had Michael Faraday's shoulders to stand on.  Now we must remember that in the 1870s, the great electromagnetic wave theory of Maxwell 
  had yet to be experimentally verified. Maxwell died young at 48 in 1879 and it is likely that had he lived longer, being then Director of the Cavendish Laboratory at Cambridge, 
  he would have initiated research into electromagnetic waves.  This task was left to a brilliant young PhD student of Hermann von Helmholtz by the name of Heinrich Hertz who published in 1887 
  the first demonstration of the existence of `invisible' electromagnetic waves as predicted by Maxwell, using the Hertz spark oscillator, now commonly known as the spark gap transmitter. 
   Hertz's studies were extensive and indeed profound, in fact his published works include studies of reflection and diffraction of radio waves (at about 70 cm wavelengths or 0.5 GHz), 
   including their transmission through material media, for example through bricks!  An astute student of science may be able to find a hint of an early demonstration of quantum mechanical 
   tunnelling in his reports \cite{Hertz1887}. We shall have more to say about this later. Tragically Hertz also died young, at just 37, his calibre as a theoretical and experimental physicist 
   was a hard act to follow, except perhaps by Enrico Fermi.  However he did not think that these `electric waves' would have any practical applications.  
   It was left to the dedication and imagination of one man, who was born in Bologna of mixed Irish and Italian heritage that gave the world modern radio technology that we have so taken 
   for granted these days.  As a matter of fact, it was exactly 124 years ago, that Guglielmo Marconi succeeded towards the end of 1901, against all criticisms from his scientific colleagues, 
   in demonstrating that he could send radio waves across the Atlantic. When news of this broke out in early 1902 the share price of the transatlantic cable companies on the London stock exchange 
   fell \cite{MarconiLeap2000}.  This was the exciting history of wireless. Marconi used a 10kW transmitter for his transatlantic experiments costing 50,000 pounds sterling in 1901 or 
   about 7.5 million in today's money.  Nowadays radio amateurs can send messages across the Atlantic with a garden dipole and a \$10 piece of electronics running merely 100mW of RF power.  In fact that is a lot of power; transmissions from Texas at a mere 1$mW$ to Western Australia is shown in Fig. 2.  That's a distance of over 16,000 kms, nearly three times that from Newfoundland to the UK with 10 million times less power than Marconi used in 1901! Marconi would be pleased. Unfortunately this is still way too high compared to the quantum limit at this frequency; more on this later.

\vspace{0.5cm}
\centerline{\includegraphics[scale=0.25]{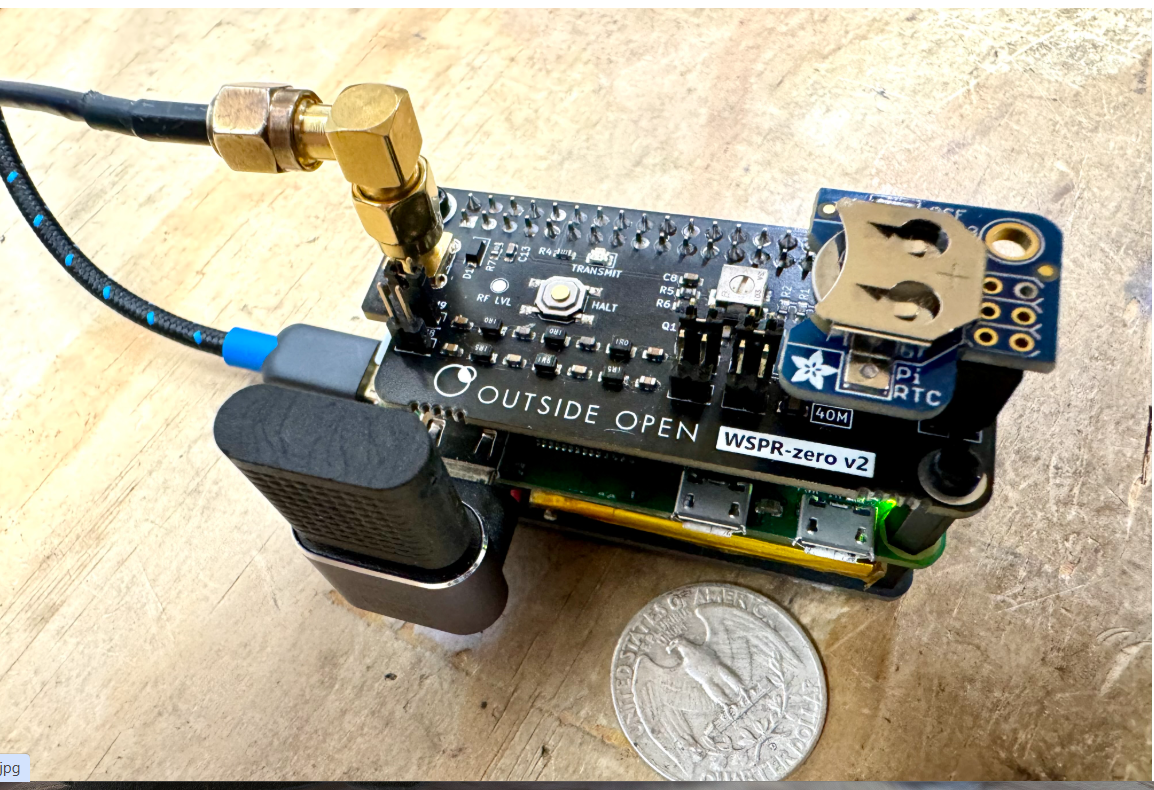}}
\vspace{0.5cm}
\centerline{Figure 1: Milliwatt WSPR transceiver built using a raspberry Pi Zero}
\vspace{0.5cm}

\vspace{0.5cm}
\centerline{\includegraphics[scale=0.45]{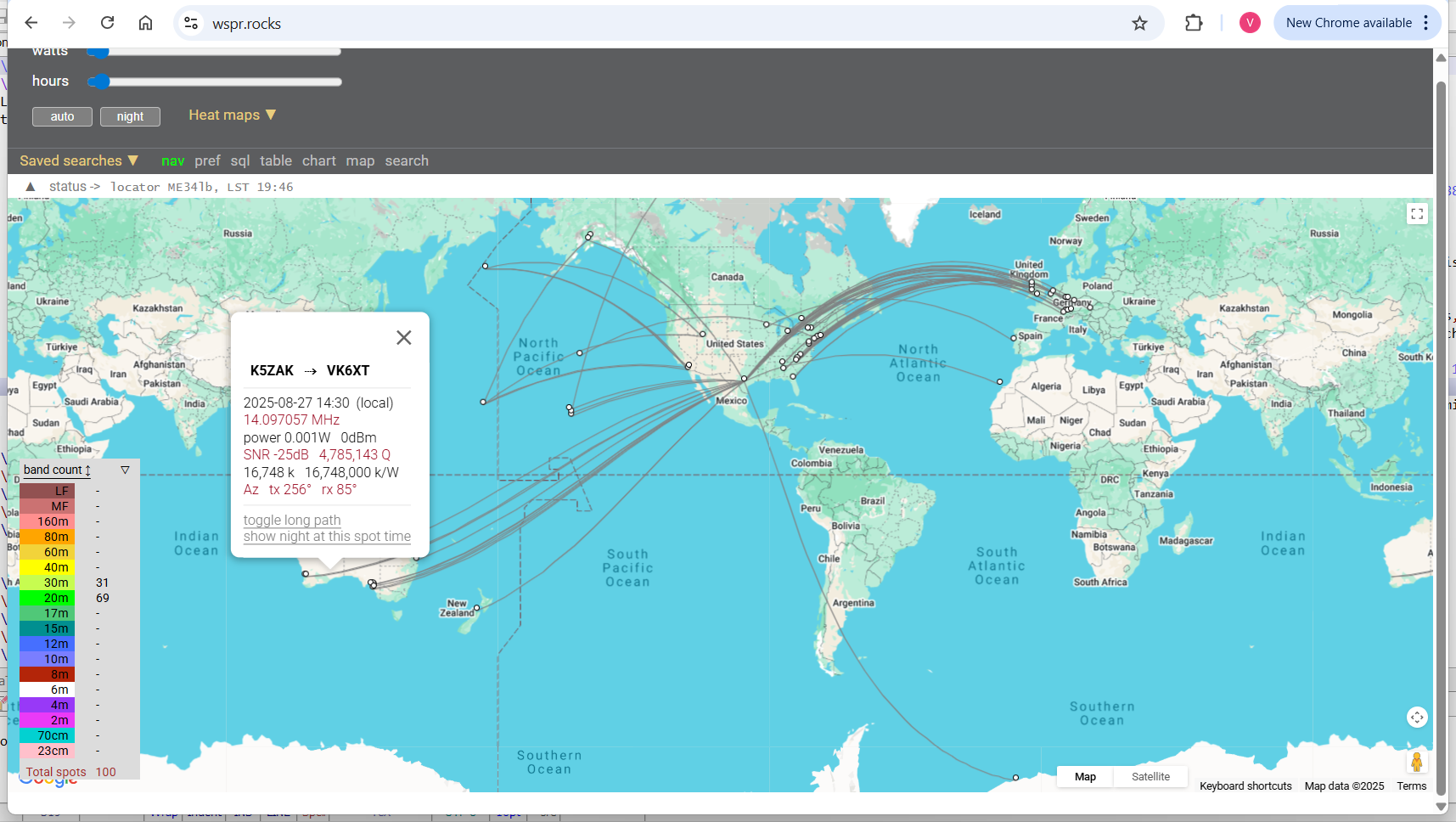}}
\vspace{0.5cm}
\centerline{Figure 2: WSPR 1mW transmissions, highlighted is transmission from Texas to Western Australia}
\vspace{0.5cm}

Back to 1901 for quietly behind in the background, Max Planck revolutionized modern physics by reluctantly introducing the light quantum hypothesis to explain the black body radiation spectrum. 
Then from 1911 to 1925, starting from the Bohr-Rutherford model of the atom;  the old quantum theory took shape and was developed by Bohr, Sommerfeld and Wilson to explain atomic spectra for more 
complicated atomic spectra.  Einstein also played a key role by introducing the photon concept for the photo-electric effect and spontaneous radiation coefficients to explain atomic radiative 
transitions, see quote above, but there were discrepancies which these theories could not explain. Then the chain of events rapidly changed when Werner Heisenberg discovered his matrix mechanics 
and within months of this discovery in 1926, in their famous three-men paper, Born, Jordan and Heisenberg made their first attempt to apply matrix mechanics to the quantization of the 
electromagnetic field according to Maxwell's theory \cite{Born-Jordan1925}. Their preliminary attempt was premature and courageous.  It took several more years before a quantum theory of light 
was finally adopted with the notable names of Heisenberg, Pauli, Dirac, Peierls, Landau and Fermi as key players.  All of these players had to involve the vector potential ${\bf A}({\bf r},t)$ 
in one way or another. Dave Delphenich has given a good account of Heisenberg and Pauli's early work at this conference \cite{DaveDelphenich}. I shall briefly fill in the gap on the others. 
Yet would Einstein agree that we have now answered his question `What are light quanta?'. I doubt it.  In this article we shall carry on the task that Basil Mahon set up in his previous 
paper \cite{BasilMahon1} by first bringing you up to date on the development of what is now widely accepted as the quantum electromagnetic vector potential ${\bf \hat A}({\bf r},t)$, with 
the hat on ${\bf \hat A}$ implying it is a quantum operator. However let's first go back to Maxwell.

At the end of Part 1 of “On Faraday’s Lines of Force” \cite{Maxwell Lines of Force} Maxwell said that he hoped, in Part 2, “to discover a method of forming a mechanical conception 
of this electro-tonic state adapted to general reasoning”. Earlier in the same paragraph he had explained that “The idea of the electro-tonic state, however, has not yet presented itself to 
my mind in such a form that  its nature and properties may be clearly explained without reference to mere symbols…”, But in Part 2 he failed to get beyond “mere symbols”. Some of the symbols, 
though, did provide a significant conceptual breakthrough (${\bf B} = curl {\bf A}$ and ${\bf E} = - d{\bf A}/dt)$. So that was the best he could do at that time. It’s a great pity that Faraday 
had lost his mental powers by the time Maxwell’s “On Physical Lines of Force” came out. Would Faraday have been satisfied by Maxwell’s representation of the electro-tonic state? We’ll never know 
for sure but Faraday said:  “Again and again the idea of an electro-tonic state  has been forced on my mind; such a state would then constitute the {\it physical lines of magnetic force}.”  
(Our italics.) Maxwell on the other hand could not have been satisfied with his mechanical model, which students are rarely taught these days, powerful though it was in predicting the
 `displacement current'. He wanted more and that took another six years before he published his `Dynamical theory of the Electromagnetic Field \cite{MaxwellDynamicalTheory}. However note the 
 prophetic words of Faraday who is known to have chosen his words carefully.  The electro-tonic `state' must be one of the most powerful and conceptually significant concepts in the history of 
 science discovered by an experimentalist with hardly any mathematical training. Maxwell on the other hand was a mathematically trained experimentalist. Few people now see him as an experimentalist 
 but he was, and in fact some of his experiments were done in his own home in London \cite{BasilMahon2003}.  He took great pains to design electromagnetic standards for measurement, he developed the 
 theory and did experiments on the viscosity of gases and opto-ellasticity and later became the first director of the now esteemed Cavendish laboratoy at Cambridge. Finally in Part III of the
  `Dynamical Theory' \cite{MaxwellDynamicalTheory} of 1864, he put his finger on Faraday's electro-tonic state as the `electromagnetic momentum' of the electromagnetic field. That  was the best 
  he could do to go beyond ``mere symbols". Unfortunately many later generation students are taught that the vector potential is not directly observable, a great error that did not come from Maxwell.
   With hindsight we can see how impressive both men were, because we had to wait until Niels Bohr who in 1911 spectacularly introduced the `stationary state' and with it the famous `quantum jumps' 
   that greatly disturbed Schr\"odinger. The quantum state then became common usage after Dirac's book in 1931. All these preoccupation with quantum states must surely come from Michael Faraday's 
   obsession with his electro-tonic state.  Subsequently Dirac identified that the Maxwell vector potential is the quantum electromagnetic potential ${\bf \hat A}({\bf r},t)$, with the hat on A that 
   actually creates or destroys these `photon' states as we are now taught.  Maxwell observed that his vector potential has a gauge degree of freedom.  He did not make much use of it.  We now know 
   that through the work of C.N. Yang and Robert Mills, Weinberg, Salam and Glashow, the gauge freedom can be generalised and with it and the idea of spontaneous symmetry breaking, the Standard Model
    with gauge symmetry group $SU(3)\times SU(2) \times U(1)$  became the pillar of modern particle physics see Yang 2014 \cite{Yang2014}.  Unfortunately, however, these advances do not bring us any 
 closer to answer Einstein's fundamental question.  In fact they complicate the situation somewhat.  The model requires the Higg's mechanism to provide mass, and for that a massive Higgs boson of 
 some $125 GeV/c^2$ had to be postulated and fortunately subsequently detected. One of the main discrepancies currently is the cosmological constant problem or vacuum energy catastrophe. 
 Extensive microwave background measurements of the universe currently place an upper bound of $2.5 \times 10^{-47} GeV^4$ in so called geometrized units $(c=1, G_m (gravitational constant) =1)$ 
 \cite{Planck Collaboration}.  Unfortunately the best current theoretical estimate based on quantum field theory is $10^8 GeV^4$, a discrepancy of some 55 orders of magnitude.  Zero point energy 
 is of fundamental significance, it prevents all atoms from collapsing into the nucleus in the ground state and is ultimately responsible for the stability of our very existence \cite{Weisskopf1985}.
 The vacuum energy catastrophe is as serious as and maybe even more serious than the ultraviolet catastrophe that had plagued classical black-body radiation theory given by the Rayleigh-Jeans law. We do not yet know the answer to such questions. Lev Landau for one never accepted quantum field theory as a logically consistent theory because of the divergent nature of the zero point energy both of the vacuum and in higher order perturbation calculations. This is not to undermine the remarkable success of the renormalization programme, as current research has shown that the Standard model is both renormalizable and mathematically consistent. However significant difficulties arise at higher energies where the supposedly graviton energy or where curved space-time issues come into play as there is as yet no unified field theory that includes gravitation. These issues become naturally important when attempting to answer cosmological questions such as the Big Bang and the origin of the universe, space-time and matter.  Elsewhere one of us has alluded to the fundamental question \cite{Choy2023} as to whether it is the quantum laws that dictate cosmology or vice-versa. If the latter, then any hope to further understand the secrets of the photon in the way Einstein desired would be a very formidable challenge indeed.

\section{Unified Quantum Electrodynamics}
\label{Unified QED}
The early pioneers of quantum mechanics were faced with a number of problems in trying to extend the finite particle quantum mechanics of Heisenberg and Schr\"odinger to a field theory with an infinite degree of freedom. Recall that in classical mechanics, we approach this by a process of coarse graining as in continuum mechanics which leads to the well known classical theories of elasticity and hydrodynamics, which are also extendable to mesoscopic systems \cite{Choy2016}. How do we do this for a quantum system, such as a quantum fluid or for the quantization of the classical electromagnetic field theory of Maxwell? We note that at the same time as others were struggling with these problems, Paul Dirac had been trying to find the relativistic wave equation for the electron, which because of the superposition principle required a first order differential equation in time. The breakthrough came when he proposed the Dirac equation in 1928 which required a new algebra, his $\gamma$ matrices later identified as a Clifford algebra  and also that the wave function must be extended to a four component object, later identified as a four component spinor. The spinor concept was first introduced by the French mathematician \'Elie Cartan in 1913 but it was rediscovered independently by Dirac in 1928. It is now well known that the triumph of this theory is the prediction of the anti-particle of the electron, namely the positron with its opposite positive charge and identical mass. However Dirac was embarrassed by the existence of negative energy states. The effort to quantize the electromagnetic field ran into the same problem. Initially there were attempts by Landau and Peierls to discover the Schr\"odinger equation for light \cite{LandauPeierls}. At the same time Heisenberg and Pauli \cite{DaveDelphenich} came up with commutation relations between the $E$ and $B$ fields which did not focus on the vector potentials. 
It became clear later, through the work of Paul Dirac and Pascual Jordan that the vector potential  is a very important object for the quantization of the electromagnetic field.  In his book of 1931 Dirac had spelled out the beautiful number operator representation of the harmonic oscillator which he could readily apply to an infinite set of non-interacting harmonic oscillators \cite{DiracBook}. As it was already well known from the classic textbook on the theory of sound by Lord Rayleigh \cite{RayleighSound}, who wrote down the theory of normal modes, Dirac immediately saw how to use normal modes to quantize Maxwell's theory using the results he had obtained for the harmonic oscillators \cite{DiracBook}.   Dirac realised that the theory has to be relativistic, so canonical quantization must be extended to a covariant form. Commutators between the four components have to be worked out which lead to the well known boson commutators for photons by identifying the Fourier coefficients of the normal modes as creation and annihilation operators. However there is a problem because quantization rules, in this case the commutators, are subjectable to the choice of gauge. Indeed there could be logical inconsistencies without further conditions which Dirac called supplementary conditions.  The first one was introduced by Enrico Fermi in (1932), which requires the choice of the Coulomb gauge (see later section \ref{Extended QED}) which is now adopted in most textbooks.  Dirac however realised that this was a major inconvenience and clearly he also saw something deeper here.  He generalised Fermi's approach to any arbitrary gauge and developed the whole method of constraint quantization to deal with it \cite{Diraclectures}.  To go into it will take us too far afield but in so doing he opened the door to constraint quantization which later in the skillful hands of Ludvig Dmitrievich Faddeev of the then USSR in the 1960s, enabled the path integral quantization of non-Abelian gauge field theories started by Yang and Mills as mentioned earlier and today also the Standard model. In his lectures of 1964 it became clear that Dirac was in fact tackling the Hamiltonian quantization of the Einstein theory of General Relativity through his method of constraint quantization.  He died without succeeding but the path he laid was so rich that it is still a major area of research; the results he left behind on weak gravity quantization for example were ground breaking.  Still the infinite zero point energy of the vacuum remains, and later the infinities in QED, which required a renormalization or other schemes to handle which we will not go into here. Instead, in this talk, we will present symbolically the first two equations of the Maxwell-Dirac electrodynamics 
which describe a fundamental feature crying out for a solution, namely the unification of the electron and photon fields. 
\begin{eqnarray}
\label{eqn1}
\frac{1}{c^2}\frac{\partial^2 A}{\partial t^2} - \nabla^2 A &=& J =\Lambda(\psi) A ,\\
(\gamma^{\mu} \partial_{\mu}  - m c)\psi &=&-A \psi.
\end{eqnarray}
The LHS of the first equation, written in the Lorentz gauge gives the wave equation for the electromagnetic potential $A(r)$ in the presence of the electron current source $J(r)=e\psi^{\dag}\gamma^\mu\psi$ which could also include an applied current if there is one on the RHS. In the second equation we have the same for the Dirac electron field $\psi(r)$ which is as mentioned earlier a spinor, while the RHS source is now the electromagnetic $A(r)$ field which may be viewed as a source.  With some skillful manipulations the first equation can also be written to look like the second with the RHS a $\Lambda$ tensor operator acting on the $A$ field. This is quite remarkable, for details see Ruei\cite{Ruei1976}.  These equations published  almost thirty years ago, gave the first author a feeling of deja vu.  One should recall it was as early as 1897 that the British mathematician and physicist Henry Cabourn Pocklington first studied 
the theory of the  current driven wire antenna \cite{Pocklington1897} that in general has a current that must be self-consistent with the electromagnetic wave it generates and vice-versa.  He solved 
this through a  series of approximations which leads to the now famous Pocklington integral equation. It was a masterpiece in classical electrodynamics.  Here we can see the same self-consistent 
phenomena at a  more fundamental level, that of Compton scattering.  Carrying this further, we can now see that the EM field and the Dirac field must be essentially two sides of the same coin. It is therefore compelling to seek a unification of both into a super symmetric field (SQED). Exactly how to do this is still an open question.  It is also not obvious that this can be done on this $(A,\psi)$ pair alone without involving the larger family of particles in the Standard model i.e. at much higher energy than the Compton wavelengths of the electron, about $10^{-11}$ cm or its classical Thomas radius $10^{-13}$ cm and at much higher energy scales than $MeV$.  However one experimental fact needs to be taken into consideration.  This is because between the low 1 MeV energy level for electron-positron annihilation with photon production or its inverse and a correspondingly very low $10^{-4}$ probability for neutrino-antineutrino production, it seems that no 
other leptons are found until the $\mu^-$ muon (at 105.7 MeV,  lifetime 2.2 $\mu s$ and the $\tau^-$ tauon (at 1.8 GeV, lifetime 0.3 ps) with their anti-particles and corresponding neutrinos.  Then there is a large empty territory for leptons all the way even until some $80-90 GeV$ for electro-weak W and Z boson production, a puzzle?    Such a unification should allow us to see what the process of electron-positron annihilation is all about. This remains a hope for the future.  We can only quote Dirac who considered the electron to be the most basic, most fundamental of all the elementary particles apart from the photon.  By great good fortune we inherit a universe in which the first and most basic, perhaps the most important lepton is stable with such a low energy scale of 0.5 MeV (lifetime estimate $\sim 10^{28}$ years), without which there will be no electronics, no radios, no televisions, no mobile phones, no computers; not to mention no chemistry, no biology and therefore we wouldn't be here! 

\section{Extended Maxwell Electrodynamics}
\label{Extended QED}
It is well known that the Maxwell-Faraday electrodynamics is locally gauge invariant. This means that all of Maxwell's equations are unchanged upon the substitution for the vector potential 
\begin{equation}
\label{eqn1b}
{\bf A}({\bf r},t) \rightarrowtail {\bf A'}({\bf r},t)= {\bf A}({\bf r},t)+ \nabla{\chi}({\bf r},t),
\end{equation}
which incidentally also requires a simultaneous change in the scalar potential $\phi({\bf r},t)$ so that:
\begin{equation}
\label{eqn1c}
\phi({\bf r},t) \rightarrowtail {\phi'}({\bf r},t)= {\phi}({\bf r},t)- \frac{1}{c} \frac{\partial\chi{\bf r},t)}{\partial t};
\end{equation}
where ${\chi}({\bf r},t)$ is an arbitrary function that defines the gauge.  The two common ones being the Coulomb gauge $\nabla \cdotp{\bf A}({\bf r},t)=0 $ and the radiation or transverse  gauge where $\nabla \cdotp{\bf A_\perp}({\bf r},t)=0$.  This is known as gauge fixing, where in the last equation ${\bf A_\perp}$ is the transverse part while the longitudinal part ${\bf A_\parallel}$ is taken to be always zero.   In recent years extensions of Maxwell's theory have been put forward for certain systems which may also require the ${\bf A}$ field itself to be gauge invariant.  For this we require a stronger condition: that is, in addition to the above gauge transformations we require the potentials to also acquire a compensating gauge field that leaves the potentials unchanged. For example in the Coulomb or radiation gauge we further demand that upon a gauge transformation:
\begin{equation}
\label{eqn1d}
{\bf A}({\bf r},t) \rightarrowtail {\bf A'}({\bf r},t)= {\bf A}({\bf r},t)+ \nabla{\chi}({\bf r},t)- \nabla{\chi'}({\bf r},t)= {\bf A}({\bf r},t).
\end{equation}
The compensating field is usually topological since $\chi$ and $\chi'$ cannot be equal but only their gradients must cancel i.e. $\nabla\chi=\nabla\chi'$. Interestingly since gauge invariance is well known to be associated with charge conservation, this entails that there must also be topological charges associated with this extension of Maxwell's theory. This issue is intricate and we won't be able to go into it here \cite{Barrett2008}.  We mention this here because other extensions of Maxwell electrodynamics have been around for quite a while, for example since Born-Infeld \cite{Born-Infeld1934} (1934) and later Dirac \cite{Dirac1962} (1962). These were introduced initially with the aim to remove divergences of the self energy but are gaining a revival of interest due to string theory and may be the key to a SQED unification as mentioned.

\section{A Poor man's QED experiments}
\label{PoormanQED}

Before moving on, let us now talk about the quantum limits and where the poor man can experiment with QED.  Here I will give you a beautiful formula from quantum electrodynamics. it comes from Landau and Lifshitz (1979) \cite{LandauLifshitzQED} and you can check it yourself.  By equating the electromagnetic energy per $m^3$ i.e. $E^2/(4\pi)$ (in Gaussian units) of an EM wave, 
with the Planck energy quanta density $\frac{h \nu}{\lambda^3} $ then the field strength of $E$ (in order of magnitude) to get into the quantum regime is given approximately by:
\begin{equation}
\label{eqn1a}
E \leq \sqrt{ 2 \hbar c}\ \frac{\nu^2}{c^2} \approx  0.25 F^2 ( \mu V/m), 
\end{equation}
where $\nu$ is the frequency in Hz while $F$ is the (dimensionless) frequency in GHz i.e. $\nu=F\times 10^9$ Hz. A static field can never be quantum \cite{LandauLifshitzQED}, 
but as it turns out, at the frequency of 1.2 GHz, your mobile phone frequency, 
the E field strength  works out as about $0.25 \mu$ V/m which is the typical sensitivity of its receivers, (or -119 dBm in rf engineer's units, assuming a 1 m antenna with 50 ohms load).  
This will give you about one photon per $m^3$ sensitivity, slightly  better than the human eye.  So a keen hacker can easily do photon counting by modifying their mobile phones and writing 
a bit of software;  many of these, including their valuable GPS receivers, are being scrapped at present.
We refer you back to our question earlier and ask our German colleagues to check if Heinrich Hertz in 1889 got anywhere close to this number in his experiments?  Photon counting should yield a Poisson distribution for such an experiment confirming the quantum nature of light. You may wonder where is a good source?  I would recommend the hydrogen 21 cm line at about 1.42 GHz , first predicted by the Dutch astronomer Frederick van de Hulst from Leiden in 1944. It is an ideal source with very stable frequency other than Doppler shifts from the Galactic spirals but you will need to pick up a satellite dish from EBay. 

Note that the quantum limit given by eqn(\ref{eqn1a}) is by no means absolute.  Consider Voyager 1 now coming up to 24.6 billion km or about one light day away from us, with its still operational 23 W transmitter.  The number of photons arriving works out with Voyager 1's 3.7m dish antenna with a gain of 47 dBi, as approximately 60 nano photon per $m^3$.  You can still receive the signals, but you will need a lot of real estate and cash for NASA's Deep Space communication systems use antennas of up to 70m diameter dishes (with about 73 dBi gain) , liquid helium cooled low noise amplifiers, very clever signal processing software and powerful computers to do the crunching. 

Now you may also wonder if your smart phone's camera can also be used for photon counting.  The answer is not quite, but there are reported experiments suitable for home experimenters for low intensity bioluminescence detection. The sensitivity is about $10^7$ photons over 180 seconds of integration time, or about $10^5$ photons per second on average \cite{HuisungKim2017}. 
One of the us (TCC) owns an  Astro T7 CCD camera for astrophotography work which is an improvement but it costs about the same price as the mobile phone. 
Suitable photon counting cameras for astronomy will costs you above \texteuro 2000, hence the GHz receiver is still a better choice if you wish to experiment.

\section{Instructive toy models}
Now it is time to take a look at some toy models for the electron. Some toy models we think are necessary even though crude, remember Maxwell's mechanical models which must have played some role in guiding him towards the correct EM theory \cite{BasilMahon1}. Lack of toy models is what might have led to many attacks on string theory. Some of the toy models we shall describe may be crude but if extended into higher dimensions may unify both the beautiful mathematics and the underlying physics of string theory.  Firstly, research into the classical inner structure of the electron started from none other than the esteemed Hendrik Lorentz (1853-1928) famous for the Lorentz transformation, the Lorentz force , the Lorentz oscillator model for dielectric dispersion, Lorentz theory of metals, the theory of the Zeeman effect etc for which he shared the Nobel prize with his student Pieter Zeeman. We are sure our colleague and friend Bernard Nienhaus can tell you a lot more about Lorentz and the Institute now named after him than we can.  Lorentz and Max Abraham (1875-1922) from G\"ottingen were competing for a theory for the inner structure of the electron using classical electrodynamics. Note that this was before Einstein's special theory of relativity and it was during these investigations that Lorentz discovered the Lorentz transformations - the power of toy models.  Both Lorentz and Abraham considered the model of a spin electron with Thomas classical radius $r_s \sim 10^{-13}$ cm and uniform charge density $e$. Later on Paul Dirac too joined in but much later and took on the problem in 1934, after he had already won the Nobel prize in 1933 for his quantum mechanical theory of the relativistic electron. They are now famous for the Abraham-Lorentz-Dirac force or the radiation reaction force, radiation damping or self-force for the electron, a problem with a long history \cite{Macdonald 2017} with many players including Pauli, Planck, Poincare, Landau etc, so we need no excuse for bringing it up again. Unfortunately, again we do not have time to discuss everything in great detail, but the interested reader we recommend the excellent book by Rohrlich \cite{Rohrlich2011} as well. Suffice to say, Abraham-Lorentz-Dirac gave a theory for the inertial mass of the electron considered as a spinning body of uniform charge $e$ and of finite dimension $r_s$. As we are here to celebrate Heisenberg, let us not forget to remind you too of his great contributions in S Matrix theory as a new concept for elementary particles, his positron theory in which he re-interpreted Dirac's equation for the electron as a `classical' field equation for a spin 1/2 particle subject to field quantization conditions. He also made ground breaking work on high energy meson production in cosmic ray showers. Cosmic rays are a poor man's ultra-high energy particle source, if you are prepared to climb mountains or fly balloons for the reward of detecting 1 TeV cosmic rays and their meson cascades, nowadays well within the reach of a dedicated physics undergraduate.  With the explosion of activities in string theory in the 80s we now wish to discuss some toy models to help you see where we can go with some crude models.  We shall introduce Professor Roger Jennison, who held the chair of physical electronics and also radio astronomy at Canterbury during the 1960s and 70s until his retirement. He founded the Electronics Laboratories there and as part of his work he had dedicated some time to researching building, theorizing and experimenting on microwave cavity models for the structure of the electron \cite{Jennison1979,Jennison 1983}.  The main idea is to capture some light and this is easy to do with microwaves e.g. from the 3K cosmic background radiation of about 3mm wavelength, see Fig 3 as originally suggested by Jennison \cite{Jennison 1983}. It is admittedly not so easy at the $\gamma$ ray wavelength of $10^{-11}$ cm for the real electron experimentally,  so here we  will have to use our imagination. 
If one could trap an EM  travelling wave say propagating in the clockwise direction, around a circular transmission line.  Now by spinning the circular cavity in the anti-clockwise direction at
 the speed of light $c$ in the medium, the wave is now effectively stationary in the laboratory frame.  Note that it is a trapped travelling wave and not a standing wave. For all intents and purpose, 
the trapped wave is now identical to a static E field and a static B field.  If the trapped light is exactly one wavelength, then the model will have a B field that could represent an electron 
with a magnetic dipole moment. In fact Jennison argued that the g factor will be exactly 2 in such a model \cite{Jennison1979}.  

\vspace{0.5cm}
\centerline{\includegraphics[scale=0.45]{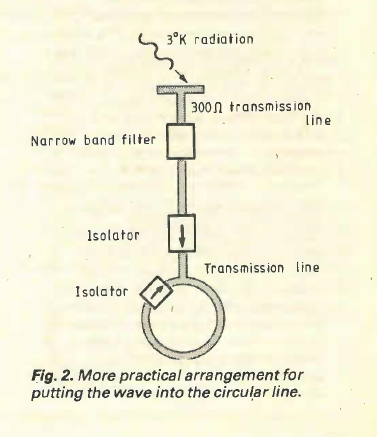}}
\vspace{0.5cm}
\centerline{Figure 3: Reproduced from Electronics World http://www.electronicsworld.co.uk with permission\cite{Jennison 1983}}
\vspace{0.5cm}

Although he did not spell it out in detail, we will quantify it as follows. The magnetic moment of such an object is given by ${\mu}=\frac{e\hbar}{2 mc} {s}$ with $s$ the spin quantum number.  Now the magnetic energy $\frac{\mu_0 B^2}{8\pi}$ is only half the total energy ${\cal E}= mc^2$ hence this is equivalent to reducing $m$ by half .  Therefore we have ${\mu}=\frac{e\hbar}{ mc} {s}$ i.e. a $g$ factor of two.  In this toy model the elementary charge is no longer fundamental but is given by the radius $r_e$ from equating the electrostatic energy $\frac{e^2}{r_e}$ to $mc^2=0.5$(MeV).  Now the radius of such an ``electron" $r_e$ is in turn given by $\frac{r_e}{\lambda_c}=2\pi \alpha$ in terms of the Compton radius $\lambda_c$ with $\alpha \approx \frac{1}{137}$, the fine structure constant. This is about the same as the Thomas radius.   Unfortunately for this model the corresponding E field is that of a charged electric dipole not quite the monopole charge of the electron, so more creative thinking is needed here. Since the model is classical it is worth noting that upon quantization, topological constraints come into play as in Dirac's famous monopole argument \cite{Diracmonopole}.  Also we are dealing with rotating frames, not quite the inertial frames of special relativity, so this picture may be over simplified.  We offer this as food for thought especially for those string theorists who might like to make contact with the low energy physics laboratory.  Jennison an engineer and astronomer had also offered an LED rotating construction that demonstrates his model now published but not so well known. It is one that is easy to construct, and, more importantly, easy to experiment with.    This is shown in Fig 4 taken from Jennsion\cite{Jennison 1983}.

\vspace{0.5cm}
\centerline{\includegraphics[scale=0.45]{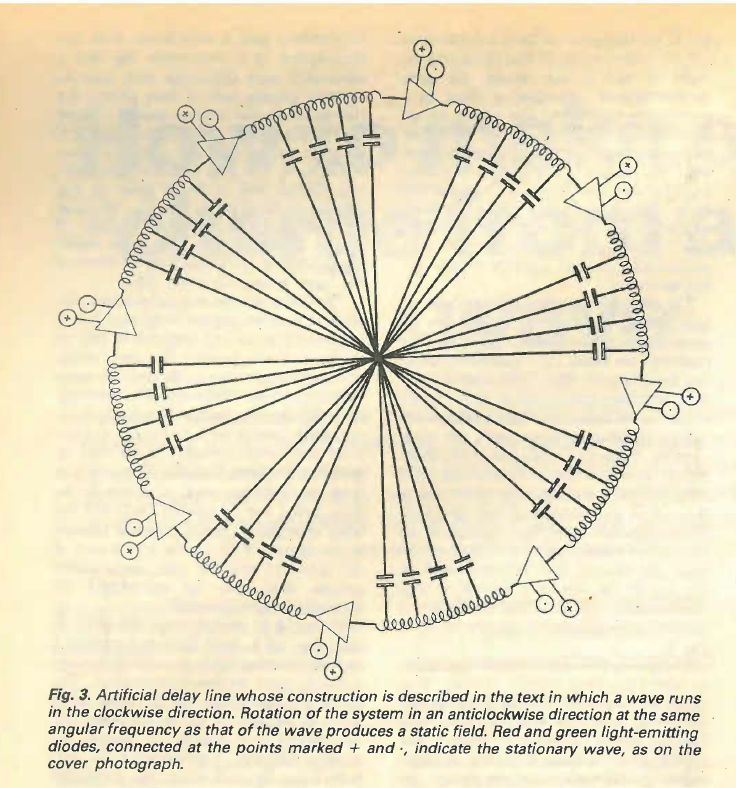}}
\vspace{0.5cm}
\centerline{Figure 4: Reproduced from Electronics World http://www.electronicsworld.co.uk with permission\cite{Jennison 1983}}
\vspace{0.5cm}

One must always remember to play with 
toy models and try to develop one's  physical intuition with them rather than start from the mathematics and be bogged down by equations.  
There was a famous account of Leon Foucault, who carried a model of the earth with a pendulum on it,  daily playing with it while he took his walk in the park in Paris in 1851, ignoring the objections of all the famous mathematicians including Cauchy, Poisson, Binet etc. Foucault with no training in mathematics, came up with his sine law for the Foucault pendulum, $(T = \frac{24{\rm hours}}{sin \theta})$, while the mathematicians were all still scratching their heads \cite{Aczel2003}. We would encourage students to build his table top model and try to refine it further. Other toy models are also available, Mark Davidson at this conference has also worked on similar models but based on line charges which we shall refer you to his work \cite{Davidson2023}. In an old unpublished preprint, one of us Choy\cite{Choy2005}, have also highlighted how useful
it would be for  physics education if students are brought to realise that the D field in Maxwell's equation and the electron charge $e$ are in fact the same, 
after all the acute student would have noticed that in SI units, they are measured in Coulombs per square metre and in Coulombs respectively.  In addition Maxwell's wisdom in insisting on writing out all his equations in full component forms \cite{BasilMahon2003} and in fact adopting the quaternion formalism clearly showed he had deep intuition about the `equivalence' of EM fields and charges \cite{Choy2005}.   

\section{A Poor Civilization's QGD experiments - Testing Quantum Gravity}
As technology progresses, we are rapidly reaching a climax in experimental physics in which quantum gravity can be tested. So let's see how we can do that using a poor civilizations's graviton detector, modelling on section \ref{PoormanQED} eqn(\ref{eqn1a}).  For gravity the analog of the E field is the strain field ${\cal H}$ or strain per unit length.  
To save you a lot of work we have written it out below, for derivation see for example Landau and Lifshitz\cite{LandauLifshitz2}:
\begin{equation}
{\cal H} \leq \frac{8\pi}{c^4}\sqrt{G\hbar c}\ \nu^2 \approx 4.6 \times 10^{-39} F^2 {\rm km}^{-1}, 
\label{PoormanGravitonEqn}
\end{equation}
where $G$ is the universal gravitational constant, $\nu$ is the frequency in Hz, while $F$ as before is the (dimensionless) frequency in GHz i.e. $\nu=F\times 10^9$ Hz.  
The complicated factors here come from the fact that gravity wave sources are quadrupoles unlike electromagnetic wave sources which are dipoles.  We can compare this to the latest Laser Interferometer Gravitational Wave Observatory (LIGO) specifications. LIGO currently has an aperture of about 4km and therefore a sensitivity to strain fields of ${\cal H} \sim 10^{-22}$/km and is designed for the 100-200Hz sweet spot to capture signals from black hole or neutron star mergers. Its proposed upgrade is LISA for Laser Interferometer Space Antenna to be placed in earth orbit.  While its proposed long baseline of 2500 km is welcome, unfortunately current technology allows it to only operate in the milli-Hertz region, the wrong direction for detecting gravitons.  Interestingly and perhaps not surprisingly, GHz LIGO observatories will require new technology in quantum metrology, super-entangled state detectors, superconducting high field magnetic levitating detectors etc to make them work at the GHz region, potentially detecting the quantum graviton signals from sources such as primordial black holes, cosmic strings and other exotic astrophysical phenomena we are yet to find out about, some possibly coming from even `before' the beginning of the Big Bang?  Young members at this conference who will live longer than most of us will be the ones privilaged to witness the discovery of the confirmation of quantum gravity as well as new astrophysical phenomena yet unknown.

\section{Conclusion}

In conclusion I would like to bring you back to 1972 just over 50 years ago at the Trieste symposium in celebration of Paul Dirac's 70th birthday and also close to the 50th anniversary of Heisenberg's quantum mechanics which we are all celebrating today. I refer to none other than our great man Heisenberg himself who gave an excellent paper entitled `Development of concepts in the history of quantum theory\cite{Heisenberg1972}'. Heisenberg  proposed that the concept of the elementary particle has to be greatly revised, he proposed the S matrix description at about this time.  We have found here that the concept of the electro-tonic state discovered by Michael Faraday, a remarkable man with little formal education, but who had a profound idea that has evolved into the U(1) gauge field of quantum electrodynamics,  and more. Now we think the gauge group of quantum gravitodynamics is the Diffeomorphism Group but this is still a matter under current research and not at all confirmed. Let us not forget that the history of physics is occasionally dotted with men like Michael Faraday, Leon Foucault, Heinrich Hertz and women like Agnes Pockels who were not household names or big guns in academia.  So we leave you with some wise words from our three great men who were founders of quantum field theory 100 years ago, words spoken during the questions and answers session after Heisenberg's talk in which there was in fact quite some heated arguments:

Werner Heisenberg:  There remains the question: What then has to replace the concept of a fundamental
particle? I think we have to replace this concept by the concept of a fundamental symmetry. 

Paul Dirac: I would like to ask Heisenberg a question. I am in general agreement with this point
of view about elementary particles that a concept really doesn't exist, but there is one reservation. I
wonder whether the electron should not be considered as an elementary particle. It may be that I am
prejudiced because I have had success with the electron and no success with other particles. I would
like to hear Heisenberg's view on that.

Werner Heisenberg: I cannot see that one could consider the electron as an elementary particle in the old
sense, because an electron can produce light quanta. Light quanta can produce baryons. So actually
the electron is connected with this world of baryons and hadrons and so on. So I don't see that you
can separate it out. As soon as an electron has these interactions, then, of course, it is also surrounded
by a cloud consisting of all these other things. Would you not agree?

Eugnene Wigner: The electron state, I believe, is surely orthogonal to the only other state with spin
one-half, namely the proton, and it is surely also orthogonal to the neutron. The neutron is surely not,
with a certain probability, just a single electron.

Werner Heisenberg: Is the electron orthogonal to helium ion which consists of a helium nucleus plus one
electron? The helium ion then transforms like the electron, and I think these two things are not 
orthogonal to each other.

Eugene  Wigner: Excuse me, I hate to argue with you, but the helium atom has four protons in it and
therefore it has a baryon number four.

Werner Heisenberg: Oh yes, my example was wrong. Now let me see...

So we start again with our quote from Albert Einstein at the very beginning of this paper, is the Photon an elementary particle?  Indeed may I also ask if the Graviton is an elementary particle?

\medskip
\noindent \textbf{Acknowledgments}
The author is indebted to Basil Mahon for his assistance in writing paragraph 3 in the introduction. This paper is dedicated to the memory of Nancy Forbes (deceased), a nuclear physicist who co-authored with Basil Mahon the top seller 'Faraday, Maxwell, and the Electromagnetic field' (How two men revolutionized physics), Promethus Books, Connecticut USA (2019).  For details of Nancy's unusual career path, see https://engage.aps.org/fgsa/resources/careers/non-academic-careers/nancy-forbes. Special thanks goes to Stella Josifovska, editor of Electronics World, UK for her permission to reuse in Figs 3 and 4 the original figures in Wireless World 1983 publication of Roger Jennison\cite{Jennison 1983}, gratis.

\end{document}